\documentclass[11pt,a4paper]{article}
\usepackage{mathrsfs}
\usepackage{cite}
\usepackage{subfigure}
\usepackage{graphicx}

\newtheorem{theorem}{Theorem}[section]

\newtheorem{lemma}[theorem]{Lemma}
\newtheorem{proposition}[theorem]{Proposition}
\newtheorem{corollary}[theorem]{Corollary}
\usepackage[cmex10]{amsmath}
\usepackage{amssymb}  
\usepackage{algorithmic}

\begin{document}
\title{On Quasi-Isometry of Threshold-Based Sampling}

\author{Bernhard A. Moser\\
 Software Competence Center Hagenberg, Austria\\
Email: bernhard.moser@scch.at }

\maketitle
\begin{center}
{\bf Abstract}
\end{center}
\noindent
The problem of isometry for threshold-based sampling such as integrate-and-fire (IF) or send-on-delta (SOD) is addressed. 
While for uniform sampling the Parseval theorem provides isometry and makes the Euclidean metric canonical, there is no analogy for threshold-based sampling.
The relaxation of the isometric postulate to quasi-isometry, however, allows the discovery of the underlying metric structure of threshold-based sampling.
This paper characterizes this metric structure making Hermann Weyl's discrepancy measure canonical for threshold-based sampling.

%

\section{Introduction}
\label{intro}
This paper addresses the question whether there are distance measures 
which are preserved by threshold-based sampling. 

Shannon's sampling theorem can be proven by means of the Parseval theorem 
showing the fundamental significance of the isometry property for a sampling theory.
The  Parseval theorem states that the Fourier transform preserves the inner product, and thus  
the Euclidean metric. To this end the Parseval theorem distinguishes the Euclidean metric in Fourier analysis and makes it canonical. 

Is there an analogy for threshold-based sampling? 
The answer is two-fold: a) no, there is no metric which provides isometry, but, b) there is a metric which is also distinguished and unique 
with respect to a relaxed version of isometry, namely quasi-isometry.
This paper provides the mathematical analysis of the distinguished role of Weyl's discrepancy measure in this context. 
Thus one could state that there is a canonical metric for threshold-based sampling which refers to the discrepancy measure.


The analysis of this paper applies to threshold-based sampling schemes which, roughly speaking, 
consist of an evaluation function $\mu(f): I\rightarrow \mathbb{R}$, for intervals $I = [t_k, t_{k+1}]$ and input signals $f$ from some 
appropriate function space and a threshold constant $\vartheta>0$ or, more general, a decreasing threshold function $\vartheta = \vartheta(t)$. 
Given a sampling point in time, $t_k$, the first occurrence of the event  $|\mu(f)[t_k, t]|\geq \vartheta(t-t_k)$, $t \geq t_k$,  
triggers the next sampling point, $t_{k+1}$. For details see \cite{Moser17}.

Though this sampling model covers various non-uniform sampling schemes such as 
Send-on-delta (SOD), level-crossing sampling with hysteresis (LC) or Integrate-and-Fire (IF), it does not 
include level-crossing without hysteresis where all crossings are considered as samples, 
level-crossing sampling with a single level, e.g. zero-crossing, sine-crossing sampling, see~\cite{Selva2012} or signal extrema sampling, 
see~\cite{Sharma2014}.

Send-on-delta (SOD) sampling is the simplest threshold-based sampling scheme. 
Given a threshold $\vartheta>0$ and an input signal $f$, SOD triggers an ``up'' or ``down'' pulse at instant $t = t_{k+1}$ depending on whether the difference $f(t) - f(t_k)$ crosses the level $f(t_k) \pm \vartheta$. 
In contrast to SOD level-crossing sampling with hysteresis (LC) w.r.t. threshold $\vartheta$ 
is based on a signal-independent lattice of levels 
$\{ k\vartheta|\, k \in \mathbb{Z}\}$, where only the first occurrence of subsequent crossings at the same level 
are taken into account the sampling shows hysteresis.  
The IF model is closely related to the SOD sampling scheme.
Instead of the difference $f(t) - f(t_k)$ the integral $\int_{t_k}^t f(s) u(s) ds$ is evaluated.
A standard choice for the averaging function $u$ is $u(t)= \exp({- \frac{t- t_k}{\alpha}})$.
In the case of $\alpha = \infty$ the IF model can be looked at as SOD, that is applied to
the function $g(t)= \int_{t_0}^t f(s)ds$. Therefore, the results regarding SOD can be directly applied to the IF model, at least for large 
$\alpha$. In biology and computational neuroscience, the IF model is well known 
in the context of modeling the functionality of a neuron~\cite{Gerstner02,Rieke1999}.
For applications as sampler in electronic systems see, e.g.,~\cite{Wei2005,ChenLXHP06}.
For a survey regarding the recovery problem see, e.g.,~\cite{LST05,Feichtinger2012}.


While for threshold-based sampling
we gain the sampled sequence of time points as an additional source of information,
we lose certain mathematical properties which are supporting pillars in the classical Nyquist-Shannon sampling theory. 

First of all, we lose linearity of the sampling operator $\Phi$.
SOD with $\vartheta=1$ applied on $f, g: [0,1] \rightarrow [0,1]$, $f(t) = \min\{1/2,t\}$, $g:= f$, illustratively demonstrates this non-linearity effect. Both signals are below threshold. Particularly, $\Phi(f)(1/2) = 0 = \Phi(g)(1/2)$.
But, the sum $f+g$ is not below threshold anymore and $\Phi(f+g)(1/2) = 1 \neq 0 = \Phi(f)(1/2) + \Phi(g)(1/2)$.

Secondly, we lose continuity w.r.t. the sampling parameter.
Uniform sampling maps a function $f: [0,T] \rightarrow \mathbb{R}$, 
$T>0$, to a sequence of function values sampled at equidistant time instants $t_k = \Delta\cdot k$, $k = 0, 1, \ldots, \lfloor T/\Delta \rfloor$, with time step parameter $\Delta>0$. 
Equipping the Cartesian product $[0,T]\times \mathbb{R}$ with the Hausdorff metric $d_H$ based on the Euclidean distance in $\mathbb{R}^2$ and assuming continuity of $f$, i.e., $f \in \mathcal{C}[0,T]$, we
obviously obtain continuity of the graph 
$\Gamma_{\Delta} := \bigcup_k \{(t_k, f(t_k))\}$ as function of $\Delta$, i.e.,
$\lim_{\Delta_n \rightarrow \Delta_0} d_H(\Gamma_{\Delta_0}, \Gamma_{\Delta_n}) = 0$.
This is no longer true in general for threshold-based sampling. 
Analogously, we obtain a graph $\Gamma_{\vartheta} := \bigcup_k \{(t_k, v_k)\}$.
For our example $f(t) = \min\{1/2,t\}$ we get $\Gamma_{\vartheta} = \{(0,0)\}$ for $\vartheta>1/2$
and $\Gamma_{1/2} = \{(0,0),(1/2, 1/2)\}$.
 Therefore, 
$\lim_{\vartheta_n \downarrow 1/2} d_H(\Gamma_{\vartheta_n}, \Gamma_{1/2}) = 1/\sqrt{2} > 
d_H(\Gamma_{1/2}, \Gamma_{1/2}) = 0$.
For the example $f: [0,T] \rightarrow [0,1/2]$, $f(t)= (\sin(t)+1)/4$, the Hausdorff metric even increases unlimited 
with $T \rightarrow \infty$.
This illustrative examples show that the Hausdorff metric is not an appropriate choice for a distance measure in the sample space
as it can lead to arbitrary large deviations for arbitrary small changes in the threshold.

Evidently, the discontinuity effect depends on the signal, the sampling operator and also on the chosen distance measure. 
Therefore, firstly, we shall propose a $\vartheta$-discontinuity measure in Section~\ref{s:impact} that 
models the effect caused by the chosen distance measure.
Secondly, we analyze under which conditions on the metrics in the input and output space
SOD becomes a mapping for which small deviations do not refer to large deviations in the respective other space. 
This quasi-isometry property was already shown in~\cite{MoserTSP2014} based on Hermann Weyl's discrepancy norm.
In this paper we deliver the missing necessity part for a full characterization of SOD as quasi-isometry in Section~\ref{s:ISO}.  
As a byproduct on the way to this result we get a characterization of equivalent discrepancy norms in 
Section~\ref{s:CharacterizationDN}. 
Before doing so, in Section~\ref{Motivation} we start with 
motivating the outlined approach based on quasi-isometry by discussing its relevance to applications.
Section~\ref{s:notation} fixes notation and recalls basic notions from metric analysis. 

\section{Motivation}
\label{Motivation}
There are fundamental differences between the familiar Shannon based uniform sampling and 
threshold-based sampling. 
This paper together with~\cite{Moser17} provides a novel mathematical unified framework based on the concept of quasi-isometry, 
which allows to analyze and quantify non-linear effects of threshold-based sampling.

As a striking difference to Shannon based uniform sampling, threshold-based sampling is characterized by an all-or-nothing law.
Either there is a triggering sampling event at a certain time or not. 
As a consequence, the sampling output no longer has to depend smoothly on changes in the input signal. 
But let us be more precise at this point: the output no longer has to depend {\it smoothly} on changes in the input signal w.r.t. 
to the standard topology of e.g. induced by the Euclidean metric. 
 
State-of-the-art signal reconstruction methods rely on assumptions or estimations of characteristics like spectral moments or 
signal bandwidth~\cite{Strohmer2006,Feichtinger2012,Selva2015,Rzepka2017,Stein2018}. Deviations or inaccurate estimates can impair the quality of the signal reconstruction.  
The proposed quasi-isometry relation is much more general. 
As such the quasi-isometry theory provides a theoretical framework for developing methods for validating the signal reconstruction error.

While~\cite{Moser17} already points out this aspect and proposes a solution based on Hermann Weyl's dicrepancy metric~\cite{Weyl1916},
this paper provides the mathematical analysis that there is no other way.
This means that Hermann Weyl's discrepancy metric is, so to speak, the {\it canonical metric} for threshold-based sampling.
To this end, this paper contributes to an understanding of threshold-based sampling and its theoretical foundations. 
The focus lies on analysis. For applications the reader is referred to~\cite{Moser17}.
 
\cite{Moser17} introduces such an alternative topology in this context. 
The starting point of this construction relies on Hermann Weyl's discrepancy measure (see,~\cite{Weyl1916,Doerr2014}). 
The resulting metrics obtained in this way turn out to be compatible with the sampling scheme in the sense that the threshold-based sampling operator becomes a quasi-isometric mapping, and that for sufficiently small thresholds, both the metrics in the input and output space become asymptotically identical. It is interesting and important that this quasi-isometry approach does not require further conditions or pre-knowledge about the input signals.

As such this framework goes beyond restrictions in terms of conditions on signal characteristics such as bandlimitedness encountered in the signal reconstruction context. 

Summing up from \cite{Moser17}, we know that there are certain metrics that satisfy the conditions of a quasi-isometry.
There may be other alternative ways to construct quasi-isometries for threshold-based sampling.
So, the question remains open whether there are other solutions than the approach based on Weyl's discrepancy measure.

This paper gives a full characterization of this problem by reference to send-on-delta sampling as simplest variant of 
a threshold-based sampling scheme.
It turns out that the choice of Weyl's discrepancy or a quasi-isometric variant is not only sufficient, 
it is also necessary in order to turn the threshold-based sampling into a quasi-isometric mapping.

\section{Mathematical Notation and Prerequisites}
\label{s:notation}
$1_{I}$ denotes the indicator function of the set $I$, i.e., $1_I(t)=1$ if $t\in I$ and $1_I(t)=0$ else.  
${\bf 0}$ denotes the constant function ${\bf 0}(t)=0$ for all $t \in \mathbb{R}$.
$\|.\|_{\infty}$ denotes the uniform norm, i.e., $\|f-g\|_{\infty} = \sup_{t \in X} |f(t)-g(t)|$, where $X$ is the domain 
of $f$ and $g$. If $M$ is a discrete set then $\#M$ denotes the number of elements. 
$a_n \nearrow a_0$ means that the monotonically increasing sequence $(a_n)_n$ converges towards $a_0$ from below.

\subsection{Mathematics of Distances}
\label{ss:distances}
In this section we recall basic notions related to distances such as semi-metric, isometry, quasi-isometry and coarse-embeddings, see e.g.,~\cite{EncyclopediaofDistances2009}.

Let $X$ be a set. A semi-metric $d:X\times X \rightarrow [0,\infty)$ is characterized by a) $d(x,x)=0$  for all $x\in X$, b) $d(x,y)=d(y,x)$ for all $x,y \in X$ and c) the triangle inequality
$d(x,z)\leq d(x,y) + d(y,z)$ for all $x,y,z \in X$. 
The semi-metric  $\tilde d$ is called {\it equivalent} to $d$, in symbols $d \sim \tilde d$, if and only if there are constants 
$A_1, A_2>0$ such that
\begin{equation}
\label{eq:norm-equivalencecondition}
A_1 d(x,y)  \leq \tilde d(x,y) \leq A_2 \, d(x,y)
\end{equation}
 for all $x$, $y$ of the universe of discourse.

A map $\Phi: X\rightarrow Y$ between a metric space ${(X,d_{X})}$ and another metric space  
$(Y,d_{Y})$ is called {\it isometry} if this mapping is distance preserving, i.e., for any 
$x_1,x_2 \in X$ we have $d_{X}(x_1,x_2) = d_{Y}(\Phi(x_1), \Phi(x_2))$.

The concept of {\it quasi-isometry}  relaxes the notion of  isometry by imposing only a 
coarse Lipschitz continuity and a coarse surjective property of the mapping. 
$\Phi$ is called a {\it quasi-isometry} from 
$(X,d_{X})$ to $(Y,d_{Y})$ if there exist constants 
$A\geq 1$, $B\geq 0$, and $C\geq 0$ such that the following two properties hold:
\begin{itemize}
\item[i)] For every two elements $x_1, x_2\in X$, the distance between their images is, up to the additive constant $B$, within a factor of $A$ of their original distance. This means,  
$\forall x_1, x_2\in X$
\begin{eqnarray}
\label{eq:quasi1}
{\frac{1}{A}}\,d_{X}(x_1,x_2)-B  &\leq & d_{Y}(\Phi(x_1),\Phi(x_2)) \nonumber\\
																 &\leq & A\,d_{X}(x_1,x_2)+B. 
\end{eqnarray}
\item[ii)] Every element of $Y$ is within the constant distance $C$ of an image point, i.e.,
\begin{equation}
\label{eq:quasi2}
\forall y \in Y:\exists x\in \mathcal{F}:d_{Y}(y,\Phi(x))\leq C.
\end{equation}
\end{itemize}

Note that for $B=0$ the condition (\ref{eq:quasi1}) reads as Lipschitz continuity condition of the operator $\Phi$.
This means that (\ref{eq:quasi1}) can be interpreted as a relaxed bi-Lipschitz condition.
The two metric spaces $ (X,d_{X})$ and $(Y,d_{Y})$ are called {\it quasi-isometric} if there exists a quasi-isometry $Q$ from $ (X,d_{X})$ to $ (Y,d_{Y})$.
A quasi-isometry with  $A=1$ is called a {\it coarse-isometry}.

As further relaxation of a metric preserving mapping Gromov introduced the concept 
of coarse-embeddings~\cite{Gromov1993}. The mapping $\Phi$ is called a {\it coarse-embedding} if there exist
non-decreasing functions $\rho_1, \rho_2: [0,\infty)\rightarrow [0,\infty)$ such that
\begin{equation}
\label{eq:coarseembedding}
\rho_1(d_{X}(x_1,x_2)) 
\leq
d_{Y}(\Phi(x_1),\Phi(x_2))
\leq 
\rho_2(d_{X}(x_1,x_2)), 
\end{equation}
for all $x_1,x_2 \in X$, and $\lim_{t\rightarrow \infty} \rho_1(t)=\infty$.
The metrics $d_{X}$
and
$d_{Y}$ are called {\it coarsely equivalent metrics} if there are non-decreasing 
functions $\kappa_1, \kappa_2: [0,\infty)\rightarrow [0,\infty)$ such that
\begin{equation}
\label{eq:cem}
d_{Y} \leq \kappa_1(d_{X})\, \, \mbox{and}
\,\,
d_{X} \leq \kappa_2(d_{Y}).
\end{equation}

\subsection{Send-on-Delta Sampling}
\label{s:threshold}
Given a continuous function $f \in \mathcal{C}[0,T]$, $T>0$, $f(0)=0$, SOD is the recursive process of detecting whether the  evaluation criterion 
$
|f(t) - f(t_k)| \geq \vartheta
$
for a given threshold $\vartheta>0$ is satisfied, where the detection at instant $t_{k+1}=t$
restarts the process by updating index $k$, i.e.,
\begin{equation}
\label{eq:trigger}
t_{k+1} = \inf\left\{t> t_k \big|\,  |f(t) - f(t_k)| \geq \vartheta \right\},
\end{equation}
and where, by definition, $t_0 = 0$. 
Let denote by $\tau_{\eta}$ the set $\{t_0, t_1, t_2, \ldots \}$ of time instants $t_k$ with $\eta(t_k)\neq 0$ together with the initial time point $t_0$.
In this context, the detection is called {\it event} which  is represented by its time instant $t_k$ and 
the sampling mode $v_k \in \{-\vartheta, \vartheta\}$ at $t_k$, $k\geq 1$. 
Therefore we encode an event by $v_k \cdot 1_{\{t_{k}\}}$, where 
$v_k := f(t_{k}) - f(t_{k-1})$. The resulting sequence of events will
 equivalently be represented as sum of its events, i.e., 
\begin{equation}
\label{eq:eta}
\eta(t):= \sum_{k\geq 0}v_k \cdot 1_{\{t_{k}\}}(t),
\end{equation}
where $\eta_0 = 0$ and $v_k \in \{-\vartheta, \vartheta\}$.
The mapping $f \mapsto \eta$ is denoted by $\Phi$ or $\Phi_{\vartheta}$ in case of emphasizing the value of the threshold $\vartheta$. The set of resulting event sequences w.r.t. $\vartheta$ is denoted by 
$\mathcal{E}_{\vartheta}[0,T] :=  \Phi_{\vartheta}(\mathcal{F}[0,T])$, where
$\mathcal{F}[0,T] := \big\{f \in \mathcal{C}[0,T]\big |\, f(0)=0 \big\}$ denotes the input space.
Note that a function (\ref{eq:eta}) is an event sequence, i.e., $\eta \in \mathcal{E}_{\vartheta}[0,T]$, if and only if $\tau_{\eta}$ has no accumulation point. In other words, this is the characterizing property that,
given $\vartheta>0$ and (\ref{eq:eta}), there is a function $f \in \mathcal{F}[0,T]$ such that 
$\Phi_{\vartheta}(f)=\eta$.
Further, note that 
\begin{equation}
\label{eq:hom}
\Phi_{\tilde \vartheta}(f) = \Phi_{\vartheta}\left(\frac{\vartheta}{\tilde \vartheta} f \right)
\end{equation}
and that $\mathcal{E}_{\vartheta}[0,T]$ is subset of the  vector space 
$\big\{
\sum_{k=1}^n a_k\eta_k \,\big|\, 
n \in \mathbb{N}, a_k \in \mathbb{R}, \eta_k \in \mathcal{E}_1[0,T] \big\}
$. Later on, we will consider metrics in $\mathcal{E}_{\vartheta}[0,T]$ which are induced by a norm defined in the 
corresponding vector space.

\begin{figure}
  \begin{center}
	      \includegraphics[width=0.65 \columnwidth]{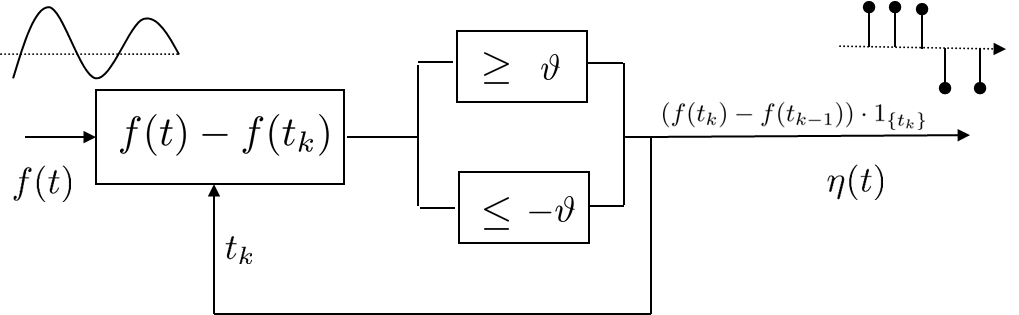} 
  \end{center}
  \caption{Block diagram of SOD according to (\ref{eq:trigger}) with input signal $f$ and
	resulting event sequence $\eta(t) := \sum_{k\geq 0} (f(t_{k+1}) - f(t_k)) \cdot 1_{\{t_{k+1}\}}$.
	}
  \label{fig:BLOCK}
\end{figure}

\begin{figure}
  \begin{center}
	      \includegraphics[width=0.65 \columnwidth]{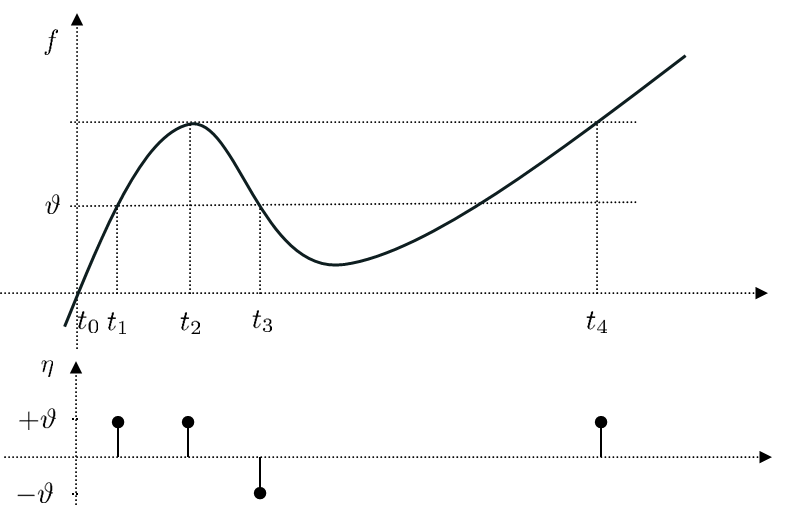} 
  \end{center}
  \caption{Example of SOD sampled signal with threshold $\vartheta$ which is equivalent to 
	level-crossing with hysteresis.
	}
  \label{fig:illustrSOD}
\end{figure}

See Fig.~\ref{fig:BLOCK} for an illustration as block diagram and Fig.~\ref{fig:illustrSOD} for an example.
Send-on-Delta (SOD) is the simplest variant of threshold-based sampling~\cite{Miskowicz2006}. 
In the  literature one can find synonymous terms and similar sampling schemes such as  
Lebesgue sampling~\cite{astrom2002}, event-based sampling~\cite{Miskowicz2007a,sanchez2009}, 
level-crossing and  level-triggered sampling~\cite{Yilmaz2012}.
Common to all these variants is that the sampling is triggered by criteria based on evaluating the signal.
Applications of the event-based sampling approach can be found in the field of wireless sensor 
networks~\cite{Miskowicz2006,Miskowicz06lontechnology,Yilmaz2013}, monitoring and control 
systems~\cite{astrom2002,Henningsson2008,Lunze2014,Ploennigs2009}, neuromorphic sensor systems~\cite{Delbruck16,Liu2014}, robotics~\cite{tobi16a,Socas2015} as well as event-based imaging~\cite{Drazen2011,tobi2016,tobi2016b}.

Note that the equivalence relation $\sim_{\vartheta}$ on $\mathcal{F}[0,T]$,
 given by 
$
f \sim_{\vartheta} g :\Leftrightarrow \Phi_{\vartheta}(f) = \Phi_{\vartheta}(g),
$
induces a one-to-one correspondence between $\Phi_{\vartheta}(\mathcal{F}[0,T])$ and the quotient space
$\mathcal{F}[0,T]|_{\sim_{\vartheta}}$. The equivalence classes 
$
[f]:= \{g \in \mathcal{F}[0,T] |\, g \sim_{\vartheta} f\}
$
 are convex subsets of $\mathcal{F}[0,T]$.  As a consequence of the discontinuity of SOD w.r.t. $\vartheta$ (see Section~\ref{s:impact}), 
these equivalence classes are neither open nor closed in the uniform topology.
Therefore, the relation between the input space and its quotient space is not obvious and needs further investigation.
The quasi-isometry approach in Section~\ref{s:ISO} will provide an answer to this question.

\section{Event Metric Discontinuity Measure (EMDM)}
\label{s:impact}
As motivated in the introduction, $\Phi_{\vartheta}$ is not continuous w.r.t. $\vartheta$. 
Nethertheless, it is left-continuous as shown next.

\begin{proposition}
\label{prop:Leftcont}
Let $\vartheta_n \nearrow \vartheta_0$ and let $\tau^{(n)} = (t_k^{(n)})_{k\in \{1, \ldots, \#\tau^{(n)}\}}$ be the sequence of sampling time points induced by the corresponding SOD sampling  operator $\Phi_{\vartheta_n}$, $n \in \mathbb{N}$ applied on $f \in \mathcal{F}[0,T]$. 
Then $t_k^{(n)} \nearrow t^{(0)}_k $ for all $k \in \{1, \ldots, \#\tau^{(0)}\}$,
$\lim_n \Phi_{\vartheta_n}(t_k^{(n)}) = \Phi_{\vartheta_0}(t_k^{(0)})$  and there is a number $N$ such that
$\#\tau^{(n)} = \#\tau^{(0)}$ for all $n \geq N$.
\end{proposition}

\noindent
Proof. 
The proof is by induction w.r.t. $k$.
Trivially, $0 = t_0 = t^{(n)}_0$ converges, i.e., $t_0^{(0)} = t_0 = 0$.
Now, suppose that $k < \#\tau^{(0)}$ and $\lim_n t^{(n)}_k = t^{(0)}_k$.

Note that $t_k^{(n)} < t_{k+1}^{(n)}$ and $t_{k}^{(n)} \leq t_{k}^{(n+1)}$ due to the construction of the sampling process (\ref{eq:trigger}) and the monotonicity assumption of $\vartheta_n$. 
Hence, $t_{k+1}^{(n+1)} \in [t_k^{(n)}, t_k^{(0)}]$.
Since $t_{k+1}^{(n+1)}$ is increasing with $n$ and bounded it converges, 
$\tilde t_{k+1} :=\lim_n t_{k+1}^{(n+1)}$. 
Let us suppose indirectly that 
\begin{equation}
\label{eq:supposetau}
\tilde t_{k+1} < t^{(0)}_{k+1}.
\end{equation}
By taking the limit w.r.t. $n$, due to the continuity of $f$ 
from
$
f(t_{k+1}^{(n)}) - f(t_{k}^{(n)}) + 
f(t_{k+1}^{(0)}) - f(t_{k+1}^{(n)}) = f(t_{k+1}^{(0)}) - f(t_{k}^{(n)}) 
$
it follows that
\begin{equation}
\label{eq:supposetau1}
f(t_{k+1}^{(0)}) - f(t_{k}^{(0)}) =  
f(\tilde t_{k+1}) - f(t_{k}^{(0)})  + f(t_{k+1}^{(0)}) - f(\tilde t_{k+1})
=  \vartheta_0.
\end{equation}
(\ref{eq:supposetau1}) together with (\ref{eq:trigger}) implies 
\begin{equation}
\label{eq:leftc1}
f(\tilde t_{k+1}) - f(t_{k}^{(0)}) < \vartheta_0.
\end{equation}
On the other hand, 
$f(t_{k+1}^{(n)}) - f(t_{k}^{(n)}) = \vartheta_n$, hence, by taking limits
we obtain $f(\tilde t_{k+1}) - f(t_{k}^{(0)} = \vartheta_0$
in contradiction to (\ref{eq:leftc1}). 
Consequently, the assumption (\ref{eq:supposetau}) cannot hold, hence 
$\lim_n t_{k+1}^{(n)} = t_{k+1}^{(0)}$ $\,\,\,\Box$

In order to quantify the discontinuity effect w.r.t. to changes in the threshold   
we propose the following measure. Since $\Phi_{\vartheta}$ is left-continuous w.r.t. $\vartheta$,
due to Proposition~\ref{prop:Leftcont}, it is sufficient to take only $\vartheta$-approximations from above into account. 

For the semi-metric $d: (\mathcal{E}_1[0,T])^2 \rightarrow [0,\infty)$ we define the Event Metric Discontinuity Measure 
(EMDM)
\begin{eqnarray}
\label{eq:stabmeasure}
{\mbox{$\Omega$}_{T}}(d) &=&  \sup_{f \in \mathcal{F}[0, T]} 
\Lambda_{f}(d), 																															 \\
\label{eq:stabmeasure1}
\Lambda_{f}(d)  &=&  
\sup_{\vartheta>0}
\limsup_{
\varepsilon \downarrow 0
}
d\left(\frac{1}{\vartheta}\Phi_{\vartheta}(f), \frac{1}{\vartheta + \varepsilon}\Phi_{\vartheta + \varepsilon}(f)\right). 	\nonumber
\end{eqnarray}
(\ref{eq:stabmeasure1}) quantifies the $d$-difference between the event sequences that result from 
arbitrarily small threshold fluctuations.

\begin{figure}
  \begin{center}
	      \includegraphics[width=0.65 \columnwidth]{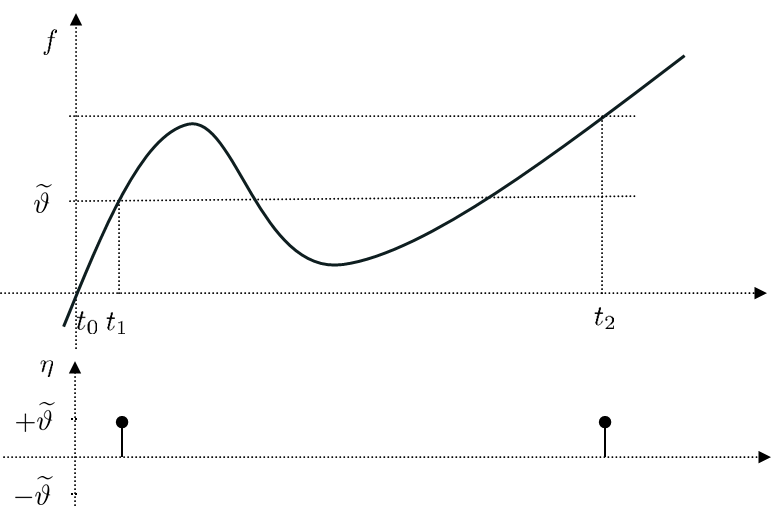} 
  \end{center}
  \caption{Example of discontinuity effect that results from approximating $\vartheta$ by 
	$\tilde \vartheta>\vartheta$ in Fig.~\ref{fig:illustrSOD}}
  \label{fig:instab}
\end{figure}

For a monotone function $f$ we obviously have  ${\mbox{$\Omega$}_T}(d) = 0$.
But, a function $f$ with local maxima or minima probably violates this condition.
See Fig.~\ref{fig:illustrSOD} and compare it with Fig.~\ref{fig:instab} for an example.
The local maximum at $t_2$ in Fig.~\ref{fig:illustrSOD} causes discontinuity (and instability in an engineering sense)
if $\vartheta$ is approximated by $\tilde \vartheta > \vartheta$ as shown in Fig.~\ref{fig:instab}.
Thus signals with a large number of local extrema, e.g., periodic signals, 
can cause massive instability effects if the event-metric $d$ is not chosen properly.

\subsection{Distances $d$ with $\mbox{$\Omega$}_T(d)< \infty$}
\label{ss:bounded}
Due to (\ref{eq:hom}), the boundedness of (\ref{eq:stabmeasure})
implies 
\begin{equation}
\label{eq:d2}
\limsup_{\lambda \uparrow 1} d\left(\frac{1}{\vartheta}\Phi_{\vartheta}(f), \frac{1}{\vartheta}\Phi_{\vartheta}(\lambda f)\right) < \infty
\end{equation}
for any $f\in \mathcal{F}[0,T]$ and $\vartheta>0$.

If there is a global bound of (\ref{eq:stabmeasure}) independent from the choice of the interval 
$[0, T]$ we say that the Event Metric Discontinuity Measure $\mbox{$\Omega$}_T(d)$ is {\it uniformly bounded}.

Let denote by 
$\mathcal{N}_{\vartheta}[0, T] = \{f \in \mathcal{F}[0,T] \, |\, \Phi_{\vartheta}(f) = {\bf 0}
\}$ 
the null-space of $\Phi_{\vartheta}$, $\overline{\mathcal{N}_{\vartheta}[0, T]}$ its closure w.r.t. $\|.\|_{\infty}$ and
$\partial \mathcal{N}_{\vartheta}[0, T] = \overline{\mathcal{N}_{\vartheta}[0, T]}\backslash \mathcal{N}_{\vartheta}[0, T]$ its boundary.

Note that $f \in \partial\mathcal{N}_{\vartheta}[0, T]$ 
implies that $\Phi_{\vartheta}(f)$  is a sequence of events of alternating signs.
Hence, by denoting the set of event sequences $\eta \in \mathcal{E}_1[0,T]$ of events with alternating signs 
by $\mathcal{E}_{\pm}[0, T]$ we obtain 
\begin{equation}
\label{eq:Epm}
\mathcal{E}_{\pm}[0,T] = \Phi_{1}\big(\partial \mathcal{N}_{1}[0, T]\big).
\end{equation}
It is interesting to point out that $\mathcal{E}_{\pm}[0,T]$ is the unit ball of 
the discrepancy norm $\|.\|_D: \mathcal{E}_1[0,T] \rightarrow [0,\infty)$, given by 
\begin{equation}
\label{eq:etaDN1}
\|\eta\|_D := \sup_{0 \leq a \leq b \leq T} \left| \sum_{t_k \in [a,b]} \eta(t_k)\right|.
\end{equation}
In the context of event sequences we denote by $d_D$ the event metric induced by $\|.\|_D$.
(\ref{eq:etaDN1}) goes back to Hermann Weyl~\cite{Weyl1916} who introduced this measure
in the context of measuring the irregularity of pseudo-random numbers.
Note that this norm is dependent on the ordering of the events which induces a highly asymmetric unit ball.
For details see~\cite{Moser2008,Moser2012a}. 

Further, note that for any $f \in \partial\mathcal{N}_{\vartheta}[0, T]$, due to the intermediate value theorem for continuous functions, we obtain 
$\lambda f \in \mathcal{N}_{\vartheta}$ for $\lambda \in (0,1)$.
The boundedness condition 
\begin{equation}
\label{eq:boundedness}
\mbox{$\Omega$}_T(d) < \infty
\end{equation}
therefore yields
\begin{eqnarray}
\label{eq:alternating1}
& & \sup_{
\tiny{\begin{array}{c}\eta \in \mathcal{E}_1[0, T]:\\
 \|\eta\|_D = 1
\end{array}}} d(\eta, { \bf 0}) = 
\sup_{\eta \in \mathcal{E}_{\pm}[0, T]} d(\eta, { \bf 0}) \\
\label{eq:alternating2}
& & =
\sup_{\vartheta>0}\sup_{f \in \partial\mathcal{N}_{\vartheta}[0, T]} d\left(\frac{1}{\vartheta}\Phi_{\vartheta}(f), {\bf 0}\right) 
\leq {\mbox{$\Omega$}_T}(d) < \infty.
\end{eqnarray}

Next we show that the inequality in (\ref{eq:alternating2}) is in fact an equality. As preliminary result we state Lemma~\ref{lemma:stabchar}.
\begin{lemma}
\label{lemma:stabchar}
Let $f \in \mathcal{F}[0,T]$ and  $\vartheta>0$. Then there are positive numbers $\vartheta^-< \vartheta$ and 
$\vartheta^+> \vartheta$ such that 
\begin{equation}
\left\| \frac{\Phi_{\vartheta_1}(f)}{\vartheta_1} - \frac{\Phi_{\vartheta_2}(f)}{\vartheta_2} \right\|_D \leq 1
\end{equation}
for all
$\vartheta_1 \in (\vartheta^-, \vartheta), \vartheta_2 \in (\vartheta, \vartheta^+)$.
\end{lemma}

\begin{figure}
  \begin{center}
	      \includegraphics[width=0.95 \columnwidth]{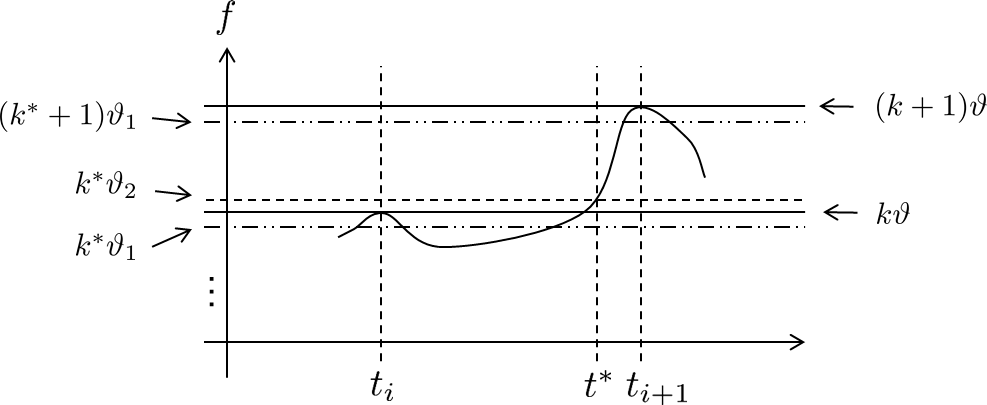} 
  \end{center}
  \caption{Illustration of (\ref{eq:t*})
	}
  \label{fig:asd1}
\end{figure}

\noindent
Proof. 
Suppose $f \neq {\bf 0}$, let $m_r = \left\lceil \|f\|_{\infty} \right\rceil\geq 1$ 
and set $\vartheta^{-} := \lambda_1 \vartheta$,  $\vartheta^{+} := \lambda_2 \vartheta$ given by 
\begin{eqnarray}
\label{eq:lambda}
\lambda_1 & = &  \frac{1}{\sqrt{1 + \frac{1}{m_r}}} < 1, \\
\lambda_2 & = & \frac{1}{\lambda_1}.
\end{eqnarray}
Note that for all $k \in \{0, \ldots, m_r-1\}$ we have
\begin{equation}
\label{eq:level}
k\, \lambda_1 < k\, \lambda_2 < (k+1)\, \lambda_1,
\end{equation}
hence, for all $\vartheta_1 \in (\vartheta^{-}, \vartheta)$, $\vartheta_2 \in (\vartheta, \vartheta^{+})$,
\begin{equation}
\label{eq:level1}
k \vartheta_1 < k \vartheta_2 < (k+1) \vartheta_1.
\end{equation}
Now suppose that 
\begin{equation}
\label{eq:qwe1}
\left\|
\frac{\Phi_{\vartheta_1}(f)}{\vartheta_1} - \frac{\Phi_{\vartheta_2}(f)}{\vartheta_2} 
\right\|_D \geq 2
\end{equation}
which implies that there is an interval $[a,b] \subseteq [0, T]$ containing at least two
 successive events of the same sign, say located at $t_i^{(\vartheta_1, \vartheta_2)}$ and $t_{i+1}^{(\vartheta_1, \vartheta_2)}$.
Now consider $\vartheta_1 \uparrow \vartheta$ and $\vartheta_2 \downarrow \vartheta$. 
The continuity of $f$ implies that the corresponding points in time are converging to some points, say
$t_i$ and $t_{i+1}$, respectively.
Without loss of generality let us assume that
\begin{eqnarray}
\label{eq:sdf1}
\eta^{-}(t_i) - \eta^{+}(t_i) & = &  1, \\
\eta^{-}(t_{i+1}) - \eta^{+}(t_{i+1}) &  = &  1, \nonumber
\end{eqnarray}
where 
$\eta^-(t) = \lim_{\vartheta_1 \uparrow \vartheta} \frac{\Phi_{\vartheta_1}(f)(t)}{\vartheta_1}$,
$\eta^+(t) = \lim_{\vartheta_2 \downarrow \vartheta} \frac{\Phi_{\vartheta_1}(f)(t)}{\vartheta_2}$.
According to Table~\ref{t:sdf1} there are $4$ possibilities to realize such an assignment.
\begin{table}
\begin{center}
\begin{tabular}{c|cccc}
   \mbox{cases}     &  $\eta^-(t_i)$  &  $\eta^+(t_i)$  &  $\eta^-(t_{i+1})$ & $\eta^+(t_{i+1})$  \\
	\hline
   1 & 1 & 0 & 1 & 0 \\
	 2 & 1 & 0 & 0 & -1 \\
	 3 & 0 & -1 & 1 & 0 \\
	 4 & 0 & -1 & 0 & -1 \\
	\hline
\end{tabular}
\end{center}
\caption{$4$ possible assignments for (\ref{eq:sdf1})
}
\label{t:sdf1}
\end{table}
Consider case $1$ of Table~\ref{t:sdf1}.
The assignment of case $1$ can  only occur  if
there are neighborhoods $U_j = (t_j - \varepsilon_j, t_j + \varepsilon_j)$ of $t_i$ and $t_{i+1}$, respectively, 
where $\varepsilon_j>0$ and $j \in \{i, i+1\}$, 
such that there is a $k^*\in \{1, \ldots, m_r\}$ with 
\begin{eqnarray}
f(t) \in ((k^*-1)\vartheta, k^* \vartheta] & \, \mbox{for all} \,  & t \in U_i, \\
f(t) \in (k^*\vartheta, (k^*+1) \vartheta] & \, \mbox{for all} \,  & t \in U_{i+1}.
\end{eqnarray}
With (\ref{eq:level1}) we obtain 
\begin{equation}
k^* \vartheta_1 < k^* \vartheta < k^* \vartheta_2 < (k^*+1) \vartheta_1 < (k^*+1) \vartheta.
\end{equation}
The intermediate value theorem for continuous functions implies the existence of
\begin{equation}
\label{eq:t*}
t^* \in \left(t_i^{(\vartheta_1, \vartheta_2)}, t_{i+1}^{(\vartheta_1, \vartheta_2)}\right)
\end{equation}
with $f(t^*) =  k^* \vartheta_2$.
See Figure~\ref{fig:asd1} for an illustration.
Note that $\Phi_{\vartheta_1}(f)(t^*)= 0$, while $\frac{1}{\vartheta_2}\Phi_{\vartheta_2}(f)(t^*) = 1$.
This means that $t_i$ and $t_{i+1}$ can not be successive points of positive events of
\begin{equation}
\label{eq:12312}
\frac{1}{\vartheta_1}\Phi_{\vartheta_1}(f) - \frac{1}{\vartheta_2}\Phi_{\vartheta_2}(f).
\end{equation} 

The other cases of Table~\ref{t:sdf1} can be treated in an analogous way by exploiting 
the intermediate value theorem. Finally we come to the conclusion that
the events of (\ref{eq:12312}) are alternating in sign, hence discrepancy bounded by $1$ $\,\,\,\Box$

\begin{theorem}
\label{th:stabchar}
 Let
 $d: (\mathcal{E}_1[0,T])^2 \rightarrow [0,\infty)$ be a semi-metric 
induced by the semi-norm $\|.\|$, i.e., $d(\eta_1, \eta_2) = \|\eta_1 - \eta_2\|$ for all $\eta_1,\eta_2 \in \mathcal{E}_1[0,T]$. Then 
\begin{equation}
\label{eq:stabchar}
\mbox{$\Omega$}_T(d) = 
\sup
\left\{
\|\eta\| \big|\, \eta \in 
\mathcal{E}_1[0,T], \|\eta\|_D = 1
\right\}.
\end{equation}
\end{theorem}

\noindent
Proof.
Due to the definition of (\ref{eq:stabmeasure}) for any
$\varepsilon >0$ there is a function $\tilde f_{\varepsilon} \in \mathcal{F}[0,T]$ and a $\vartheta>0$
with 
\begin{equation}
\label{eq:ksdfl}
\Lambda_{\vartheta, \tilde f_{\varepsilon}}(d) > \mbox{$\Omega$}_T(d) - \varepsilon,
\end{equation}
where 
\[
\Lambda_{\vartheta, \tilde f_{\varepsilon}}(d)  = 
\limsup_{
\varepsilon \downarrow 0
}
d\left(\frac{1}{\vartheta}\Phi_{\vartheta}(f), \frac{1}{\vartheta + \varepsilon}\Phi_{\vartheta + \varepsilon}(f)\right) 
\] denotes the term after the supremum of (\ref{eq:stabmeasure1}).
With Lemma~\ref{lemma:stabchar} we conclude that there are thresholds
$\vartheta^- \in (0, \vartheta)$ and $\vartheta^+ \in (\vartheta, 2\vartheta)$ such that 
\begin{equation}
\label{eq:bnjm}
\left\| 
\frac{1}{\vartheta_1}\Phi_{\vartheta_1}(\tilde f_{\varepsilon}) -
\frac{1}{\vartheta_2}\Phi_{\vartheta_2}(\tilde f_{\varepsilon})
\right\|_D \leq 1
\end{equation}
for all 
$\vartheta_1 \in (\vartheta^-, \vartheta)$ and  
$\vartheta_2 \in ( \vartheta, \vartheta^+)$, hence
\begin{eqnarray}
\label{eq:adf} 
 &      &  \sup\left\{\|\eta\| \big|\, \eta \in \mathbb{E}, \|\eta\|_D=1 \right\} \nonumber \\
 & \geq &  \limsup_{\vartheta^- \uparrow \vartheta, \vartheta^+ \downarrow \vartheta}\left\|
\frac{1}{\vartheta_1}\Phi_{\vartheta_1}(\tilde f_{\varepsilon})  -
\frac{1}{\vartheta_2}\Phi_{\vartheta_2}(\tilde f_{\varepsilon})
\right\|
\nonumber\\
& = &
\Lambda_{\vartheta,\tilde f_{\varepsilon}}(d).
\end{eqnarray}
Together with (\ref{eq:ksdfl}) and (\ref{eq:alternating2}), this proves (\ref{eq:stabchar})
$\,\,\,\Box$

Theorem~\ref{th:stabchar} tells us that (\ref{eq:stabmeasure}) is uniformly bounded for  event metrics that are equivalent to Weyl's discrepancy norm.

\subsection{Examples}
\label{ss:examples}
Theorem \ref{th:stabchar} provides a sufficient criterion that 
$\mbox{$\Omega$}_T(d)$ is bounded if the event metric $d$ is coarsely equivalent to Weyl's discrepancy norm.
Next we list examples of equivalent metrics and counter examples.

\subsubsection{Alexiewicz norm}
The Alexiewicz metric is a special case of the discrepancy measure by restricting the set of intervals 
$[a,b] \subseteq [0, T]$ to those of the form $[0, t]$, where $t \in [0, T]$, i.e.,
\begin{equation}
\label{eq:Alex}
\|\eta\|_A:= \sup_{a \in [0, T]}\left|\sum_{t_k \in [0, a], t_k \in \tau_{\eta} } \eta(t_k)\right|
\end{equation}
for $\eta = \eta_1 - \eta_2$, $\eta_i \in \mathcal{E}_{\vartheta}$, $i=1,2$, where $\tau_{\eta} = \tau_{\eta_1}\cap \tau_{\eta_2}$.
This norm was introduced by Alexiewicz~\cite{Alexiewicz1948} in the context of alternative integration concepts such as Henstock-Kurzweil integral, see also~\cite{swartz1998}. 

As shown in \cite{Moser2012a} we have
\begin{eqnarray}
\label{eq:diam}
\|\eta\|_D  &=& \max\left\{\max_{a\in [0,T]}\sum_{t_k \in \tau_{\eta}, t_k \leq a} \eta(t_k),0\right\} \nonumber \\
						&  & -\min\left\{\min_{a\in [0,T]}\sum_{t_k\in \tau_{\eta},t_k \leq a} \eta(t_k),0\right\},
\end{eqnarray}
 hence
$1/2 \|\eta\|_D \leq \|\eta\|_A \leq \|\eta\|_D$, i.e., 
(\ref{eq:Alex}) is equivalent to (\ref{eq:etaDN1}), thus (\ref{eq:Alex}) satisfies (\ref{eq:boundedness}).
Like (\ref{eq:etaDN1}) $\mbox{$\Omega$}_T(\|.\|_A)$ is uniformly bounded.

\subsubsection{Max-Max-Sum Norm}
\label{ss:mixed}
The following {\it Max-Max-Sum norm}, $\|.\|_{M}$, will serve as an example in 
Section~\ref{s:CharacterizationDN} 
to show the independence of the proposed axioms for characterizing  equivalent discrepancy (semi)-norms:
\begin{equation}
\label{eq:M}
\|\eta\|_{M} := \max\{ \max_{t_k \in \tau_{\eta}}\{|\eta(t_k)|\}, |\sum_{t_k \in \tau_{\eta}} \eta(t_k)|\},
\end{equation}
where $\eta = \eta_1 - \eta_2$, $\eta_i \in \mathcal{E}_{\vartheta}$, $i = 1,2$.
The unit ball of (\ref{eq:M}) is the intersection of the $\|.\|_{\infty}$-ball and  
the half spaces  
\begin{eqnarray}
\mathcal{H}^- &:=& \{\eta \in \mathcal{E}_{\vartheta}[0,T]\big|\, \sum_{t_k \in \tau_{\eta}} \eta(t_k) \leq 1\}, \nonumber\\
\mathcal{H^+} &:=& \{\eta \in \mathcal{E}_{\vartheta}[0,T]\big|\, \sum_{t_k \in \tau_{\eta}} \eta(t_k) \geq -1\}. \nonumber
\end{eqnarray}
Note that $\mbox{$\Omega$}_T(\|.\|_M)$ is uniformly bounded though it is not coarsely equivalent to $\|.\|_D$. For example, consider the sequence 
\begin{equation}
\label{eq:MMSN}
\eta_n(t) = \sum_{k=1}^{\lceil n/2 \rceil} 1_{\{\frac{k}{n}\}}(t) 
- \sum_{k=\lceil n/2\rceil +1}^{n} 1_{\{\frac{k}{n}\}}(t),
\end{equation}
which yields $\|\eta_n\|_{M} = 1$ for all $n\in \mathbb{N}$, while $\lim_n \|\eta_n\|_D =\infty$.

\subsubsection{van Rossum Spike Metric and Schreiber Spike Similarity}
\label{ss:rossum}
The {\it van Rossum metric} is widely used in neural computation for discriminating spike trains~\cite{Rossum01}.
This distance relies on the idea to represent an event (spike),
$\eta(t_k) 1_{\{t_k\}}(t)$, by its convolution with a causal exponential, i.e.,
\[
\rho_k(t) = \eta(t_k) e^{-\alpha (t-t_k)} 1_{[t_k,\infty)}(t)
\] 
with time constant $\alpha = 1/t_c \geq 0$ and $\eta(t_k) \in \{-\vartheta, \vartheta\}$.
By this approach, the event sequence 
$\eta(t) = \sum_{t_k \in \tau_{\eta}} \eta(t_k)1_{\{t_k\}}(t)$
is mapped to the function
$R_{\eta}(t) = \sum_{t_k \in \tau_{\eta}} \rho_k(t)$, which ``fills the gap between'' the events and allows to apply 
the familiar Euclidean metric in a meaningful way, i.e.,   
\begin{equation}
\label{eq:VR}
d_{R, \alpha}(\eta_1,\eta_2) :=  \|R_{\eta_1} - R_{\eta_2}\|_2.
\end{equation}
For an event sequence $\eta$ of alternating signs we obtain $-\vartheta \leq R_{\eta}(t) \leq \vartheta$, which shows that
the boundedness condition (\ref{eq:boundedness}) is satisfied for a compact interval $[0,T]$.
However, consider a constant $\Delta>0$ and $t_k = k \cdot \Delta$ with 
$\eta(t_k) = (-1)^k$.
By induction we get $|R_{\eta}(t_k)| \geq \sum_{m=1}^k (-1)^{m+1} e^{-m\alpha \Delta}$.
Since $|R_{\eta}|$ is monotonically decreasing on $[t_k, t_{k+1})$ we obtain
\begin{equation}
\label{eq:vRbelow}
|R_{\eta}(t)| \geq \vartheta e^{-\alpha \Delta} (1- e^{-\alpha \Delta})
\end{equation}
for all $t\in [0,T]$ and $\alpha \in (0,\infty)$. 
By this we obtain lower and upper bounds for $\mbox{$\Omega$}_T(d_{R, \alpha})$, 
\begin{equation}
\label{eq:EMDMvR}
\kappa_{\alpha,\Delta} \cdot T \leq \mbox{$\Omega$}_T(d_{R,\alpha}) \leq T,
\end{equation}
where $\kappa_{\alpha,\Delta} = e^{-2\alpha \Delta} (1- e^{-\alpha \Delta})^2 >0$ in case of $\alpha \in (0,\infty)$ and 
$\kappa_{\Delta} = 1/2$ in case of  $\alpha = 0$.
In contrary to Weyl's dicrepancy norm, (\ref{eq:etaDN1}), and the Alexiewicz norm, (\ref{eq:Alex}),  for $\alpha \in [0, \infty)$
(\ref{eq:EMDMvR}) implies
\begin{equation}
\label{eq:vRunstable}
\lim_{T\rightarrow \infty} \mbox{$\Omega$}_T(d_{R,\alpha}) = \infty.
\end{equation}
The van Rossum metric is induced by the familiar inner product 
$d_{R,\alpha}(\eta_1, \eta_2) = \langle s_{\eta_1 - \eta_2}, s_{\eta_1 - \eta_2} \rangle$,
given by the difference of convolutions
\begin{equation}
\label{eq:innerP}
s_{\eta_1 - \eta_2} := (\Delta_{\eta_1}  - \Delta_{\eta_2}) \ast \chi,
\end{equation}
where $\Delta_{\eta}$ is the comb of Dirac impulses, which  results from $\eta$ by
replacing the events $\eta(t_k) 1_{\{t_k\}}(t)$ by $\eta(t_k) \delta_{\{t_k\}}(t)$, and 
the convolution is applied on the causal exponential $\chi$.
Though widely used in applications, (\ref{eq:vRunstable}) in combination with 
(\ref{eq:stabmeasure}) means that the van Rossum metric and Schreiber similarity measures are prone to cause instabilities, at least in the context of threshold-based sampling. This effect  becomes more and more apparent with increasing time range $T$. 

Like the van Rossum metric the {\it Schreiber spike similarity measure} $S: (\mathcal{E}_1[0,T])^2 \rightarrow [-1,1]$, see, e.g.,~\cite{Dauwels2009}, 
is also defined via an inner product $\langle.,.\rangle$ and a convolution with some kernel $0 \leq \chi \leq 1$, i.e.,
\begin{equation}
\label{eq:Schreiber}
S(\eta_1, \eta_2) := \frac{\langle\eta_1\ast \chi, \eta_2\ast \chi\rangle}
{{\sqrt{\langle\eta_1\ast \chi, \eta_1\ast \chi\rangle}}{\sqrt{\langle\eta_2\ast \chi, \eta_2\ast \chi\rangle}}}.
\end{equation}
Consider $\eta_1 \in \mathcal{E}_1[0,1]$ given by $\eta_1(k/n) =  +1_{\{k/n\}}$ for $k\leq n/2$ and $\eta_1(k/n) =  -1_{\{k/n\}}$  for 
$k\geq n/2$, $\eta_2 = -\eta_1$, on the one hand, and $\eta_3$ given by $\eta_3(k/n) = (-1)^k 1_{\{k/n\}}$ and $\eta_4 = -\eta_3$.
Consider $S(\eta_1, \eta_2)= -1 = S(\eta_3, \eta_4)$. Let $h:[-1,1] \rightarrow [0,\infty]$, $h(1)=0$ be a decreasing function, in order to 
model a distance measure  based on (\ref{eq:Schreiber}), that is $d_S(\eta, \widetilde \eta) :=  h(S(\eta, \widetilde \eta))$.
Observe that the postulate $\Omega_T(d_S) \leq M <\infty$ implies 
\begin{equation}
\label{eq:SchreiberM}
d_S(\eta_1, \eta_2) = d_S(\eta_3, \eta_4) \leq M < \infty.
\end{equation}

\subsubsection{Victor-Purpura Editing Distance for Spike Trains}
\label{ss:victor}
The {\it Victor-Purpura metric} $d_{\mbox{\tiny VP},s}(\eta_1, \eta_2)$ with parameter $s>0$ is an editing distance~\cite{Victor1996}.
The editing process is realized by recursive application of either {\it deletion}, {\it insertion} or {\it shift}.
Each editing operation is associated a cost function $c$. The cost for a deletion or an insertion of a spike is given by $c=1$. The cost for a shift over $\Delta t$ is given by $c(\Delta t) := s\, \Delta t$. 
The distance between $\eta_1$ and $\eta_2$ ($\eta_i\in \mathcal{E}_1[0,T]$ and $\eta_i\geq 0$) is defined as the minimal cost of an editing process that
transforms $\eta_1$ into $\eta_2$.
 
In order to extend the idea to event sequences with $-1$ and $1$ elements we propose 
to represent an event sequence $\eta \in \mathcal{E}_1[0,T]$ by its positive and negative part, that is
$\eta = \eta^+ - \eta^-$, where $\eta^+(t) = \max\{0, \eta(t)\}$ and  $\eta^-(t) = \min\{0, \eta(t)\}$.
The editing distance between $\eta_1 = \eta^+_1 - \eta^-_1$ and $\eta_2 = \eta^+_2 - \eta^-_2$ can then be computed by
applying the editing operations on the sequences $\eta^+_1 + \eta^-_2 \geq 0$ and 
$\eta^-_1 + \eta^+_2 \geq 0$. The cost for transforming an event sequence of alternating signs at equidistant instants, 
$\eta(t) = \sum_k (-1)^k 1_{\{ \Delta*k\}}(t)$, $\eta \in  \mathcal{E}_{\pm}[0,T]$, to zero is therefore bounded from below by 
$\min\{\#\eta-1, s\, (T/2-2 \Delta)\}$. For positive shifting cost parameter $s>0$, therefore, for sufficiently small $\Delta>0$ the shift operation is less costly then insertion and deletion. By this, we obtain
\begin{equation}
\label{eq:VP}
\min\{\#\eta-1, s\, (T/2-2 \Delta)\} \leq \Omega_T(d_{\mbox{\tiny VP},s}) \leq T.
\end{equation}
Like for the van Rossum metric the EMDM maesure for Victor-Purpura increases with $T$, i.e.,
$\lim_{T \rightarrow \infty} \Omega_T(d_{\mbox{\tiny VP},s}) = \infty$.

Note that for $s=0$ the distance $d_{\mbox{\tiny VP},s}(\eta_1, \eta_2)$ reduces to a counting measure, i.e.,
$d_{\mbox{\tiny VP},s}(\eta_1, \eta_2) = |\#\eta_1^+ - \#\eta_2^+| + |\#\eta_1^- - \#\eta_2^-|$, hence
$\Omega_T(d_{\mbox{\tiny VP},s}) \leq 1$. 

\section{Characterization of Equivalent Discrepancy Norms}
\label{s:CharacterizationDN}
In this section, first of all we will exploit discrete mathematical properties of the discrepancy norm $\|.\|_D$.  
Based on the concept of minimal length intervals of maximal discrepancy (MMD intervals), introduced in~\cite{Moser2014Random}, 
we will show Proposition~\ref{prop:chain}. 
\begin{proposition}
\label{prop:chain}
Let $\eta \in \mathcal{E}_1[0, T]$, $T>0$, and $r = \|\eta\|_D < \infty$,  then there are 
$r$ many event sequences $\eta_k \in \mathcal{E}_1[0, T]$, $\eta_0 = {\bf 0}$, $\eta_r = \eta$, $k \in \{0,\ldots, r\}$ 
such that 
\begin{equation}
\label{eq:chain}
\|\eta\|_D = \sum_{k=1}^{r} \| \eta_{k} - \eta_{k-1}\|_D,
\end{equation}
where $\| \eta_{k} - \eta_{k-1}\|_D = 1$ for all $k \in \{0,\ldots, r\}$.
\end{proposition}

\noindent
Proof. 
Let $\eta \in  \mathcal{E}_1[0, T]$, $\eta \neq {\bf 0}$. Then $r = \|\eta\|_D < \infty$.
Let us recursively define the following sequence of intervals $\mbox{MMD}_{\eta,m} = [a_m, b_m] \subseteq [0,T]$, 
$b_m < a_{m+1}$, $m \in \{1,\ldots, M\}$, $M\in \mathbb{N}$: 
\begin{eqnarray}
\label{eq:MMD}
a_0 & := & 0, \\
b_0 & := & T, \nonumber\\
b_{m+1} & := & \min\left\{t_k \in \tau_{\eta} \,\left|\, \left\|\eta\big|_{[a_m,t_k]}\right\|_D = r \right.\right\}, \nonumber \\
a_{m+1} & := & \max\left\{t_k \in \tau_{\eta} \,\left|\, \left\|\eta\big|_{[t_k,b_{m+1}]}\right\|_D = r, t_k\leq b_{m+1}  \right.\right\}. \nonumber
\end{eqnarray}
Note that $\left\|\eta\big |_{\mbox{\small{MMD}}_{\eta,m}}\right\|_D = \|\eta\|_D$ and that the sequence of partial sums
\begin{equation}
\label{eq:Dsum}
D_m = \sum_{t_k \in \mbox{\small{MMD}}_{\eta,m}} \eta(t_k)
\end{equation}
 is alternating in sign. As a consequence
to this Chebyshev alternating sign property the partial sums on the in-between intervals are vanishing, i.e.,
\[
\sum_{t_k \in [b_{m}, a_{m+1}]} \eta(t_k) = 0\]
 for $m \in \{1, \ldots, M-1\}$. 

Now, set $\eta_r := \eta$ and define $\eta_k$ recursively.
Given $\eta_k$ and let denote its MMD intervals by $\{\mbox{MMD}_{\eta_k,1}, \ldots, \mbox{MMD}_{\eta_k,M_k}\}$,
$\mbox{MMD}_{\eta_k,m} = [a_m^k, b_m^k]$ and define $\eta_{k-1}$ by setting
\begin{equation}
\label{eq:rec1}
\eta_{k-1}(a_m^k) = 0
\end{equation}
for all $m \in \{1, \ldots, M_k\}$
and 
\begin{equation}
\label{eq:rec2}
\eta_{k-1}(t) = \eta_{k}(t) 
\end{equation}
on $[0,T] \backslash \{a_1^k, \ldots, a_{M_k}^k\}$.
By setting the first element of each MMD interval of $\eta_{k}$ to zero the resulting event sequence $\eta_{k-1}$ reduces its range in the walking graph by $1$, that is, its discrepancy reduces by $1$, 
$\|\eta_{k-1}\|_D = \|\eta_{k}\|_D-1$.
Since the partial sums (\ref{eq:Dsum}) are maximal and alternating in sign, the event sequence given by
\[
\eta_k - \eta_{k-1} = \sum_{m=1}^{M_k} \eta_{k}(a_m^k)
\]
 is of alternating sign. 
Consequently, $\|\eta_k - \eta_{k-1} \|_D = 1$. After $r-1 = \|\eta\|_D-1$ steps the resulting 
event sequence $\eta_1$ is of alternating signs. Each event $\eta_1(t_m) 1_{\{t_m\}}$ represents an 
MMD interval, $\mbox{MMD}_{\eta_1,m} = \{t_m\}$. Due to the recursion, (\ref{eq:rec1}) and (\ref{eq:rec2}),
we in the final step obtain $\eta_r = {\bf 0}$ $\,\,\,\Box$

In combination with Theorem~\ref{th:stabchar} the boundedness of $\mbox{$\Omega$}_T(\|.\|)$ w.r.t. a given (semi-)norm $\|.\|$,
i.e.,
\begin{equation}
\label{eq:ALT}
\sup_{\eta \in \mathcal{E}_{\pm}[0,T]} \|\eta\| < \infty,
\end{equation} 
Proposition~\ref{prop:chain}
implies
\begin{equation}
\label{eq:equ1}
\|\eta\| =   \left\|\sum_{k=1}^r \eta_{k}- \eta_{k-1} \right\|  \leq  \sum_{k=1}^r \|\eta_{k}- \eta_{k-1}\| 
				 \leq  \|\eta\|_D \, \mbox{$\Omega$}_T(\|.\|). 
\end{equation}
(\ref{eq:equ1}) refers to the right hand side inequality of the norm equivalence condition (\ref{eq:norm-equivalencecondition}).

Next we consider the condition
\begin{equation}
\label{eq:infgeq0}
\inf\left\{ \left.\frac{1}{\#\tau_{\eta}} \|\eta\| \right|\, \eta \geq {\bf 0}, \eta \in \mathcal{E}_{\vartheta}[0,T]\backslash{\{\bf 0\}}
      \right\}>0,
\end{equation}
which, of course, is not sufficient to ensure 
the left hand side inequality of (\ref{eq:norm-equivalencecondition}).

For this purpose we take up the notion of a transcription operator $T_p$ introduced in
\cite{MoserEBCCSP15a} which replaces patterns of the form 
$p = (+1, 0, \ldots, 0, -1)$ in the sequence $\eta$ by a $0$-sequence of the same length.
In addition to $(+1, 0, \ldots, 0, -1)$ patterns we also consider $(-1, 0, \ldots, 0, +1)$ patterns and refer to the 
corresponding operators as $T_{(+-)}$ and $T_{(-+)}$, respectively.
$T_p$ can be applied recursively, i.e., $T_p^{n+1}(\eta) = T_p(T_p^{n}\eta)$ and $T_p^0(\eta) =\eta$.
As $T_{(+-)}^n$ makes shortcuts in the walking graph $\Gamma = \{ (t_k, \sum_{m=1}^k \eta(t_m))_k\, |\, t_k \in \tau_{\eta}\}$ 
and $\|.\|_D$ equals the range of $\Gamma$, see (\ref{eq:diam}), we obtain the inequality 
\begin{equation}
\label{eq:Tinequ}
\|T^n_p(\eta)\|_D \leq \|\eta\|_D
\end{equation}
for arbitrary $n \in \mathbb{N}$ and $p = (+-)$ or $p=(-+)$.
(\ref{eq:Tinequ}) implies that there is a constant $c>$ such that
\begin{equation}
\label{eq:Tinequ1}
\max_{n,m,I}\left\|T^n_{(-+)}\big(T^m_{(+-)}(\eta\big|_I)\big) \right\| \leq c \|\eta\|
\end{equation}
for an equivalent discrepancy norm, $\|.\| \sim \|.\|_D$, for all $\eta \in \mathcal{E}_1[0, T]$ and all intervals $I \subseteq [0, T]$.
Note that (\ref{eq:Tinequ}) is not satisfied for $\|.\|_M$, see (\ref{eq:M}).

(\ref{eq:Tinequ1}) turns out to be a characteristic property of $\|.\|_D$ as stated in Theorem~\ref{th:equDN}.
\begin{theorem}
\label{th:equDN}
 Let $d: (\mathcal{E}_1[0,T])^2 \rightarrow [0,\infty)$ be a semi-metric 
induced by the semi-norm $\|.\|$, i.e., $d(\eta_1, \eta_2) = \|\eta_1 - \eta_2\|$ for all $\eta_1,\eta_2 \in \mathcal{E}_1[0,T]$, and let
$d_D: (\mathcal{E}_1[0,T])^2 \rightarrow [0,\infty)$ bet the metric induced by the discrepancy norm $\|.\|_D$.
Then, $d$ is equivalent to $d_D$ if and only if $\|.\|$ satisfies (\ref{eq:ALT}), (\ref{eq:infgeq0}) and (\ref{eq:Tinequ1}).
\end{theorem}

\noindent
Proof.
The necessity of (\ref{eq:ALT}), (\ref{eq:infgeq0}) and (\ref{eq:Tinequ1}) given $d \sim d_D$ is a direct consequence of (\ref{eq:norm-equivalencecondition}).
Now, let us assume (\ref{eq:ALT}), (\ref{eq:infgeq0}) and (\ref{eq:Tinequ1}).
Due to Proposition~\ref{prop:chain} we obtain the right hand side of  (\ref{eq:norm-equivalencecondition}), i.e.,
$\|\eta\|\leq A \|\eta\|_D$, where $A=  \sup_{\eta \in \mathcal{E}_{\pm}[0,T]} \|\eta\| < \infty$.

Let  $\eta \in \mathcal{E}_1[0,T]\backslash {\{\bf 0\}}$ with $r = \|\eta\|_D$. 
Consider the first MMD interval $[a_1,b_1]$ due to (\ref{eq:MMD})
which defines the
mapping $\Pi: \mathcal{E}_1[0,T] \rightarrow \mathcal{E}_1[0,T]$,
\begin{equation}
\label{eq:Pi}
\Pi(\eta) := T^r_{(-+)} \big( T^r_{(+-)} (\eta|_{[a_1,b_1]}) \big).
\end{equation} 
Note that $\|\Pi(\eta)\|_D = \|\eta\|_D$ and that $\Pi(\eta)$ is an event sequence where all events have the same sign.
Consequently, by (\ref{eq:infgeq0}) and (\ref{eq:Tinequ1}), we obtain
\begin{equation}
c \inf_{\eta \neq {\bf 0}} \frac{\|\eta\|}{\|\eta\|_D} 
\geq  \inf_{\eta \neq {\bf 0}} \frac{\|\Pi(\eta)\|}{\|\Pi(\eta)\|_D} 
=  \inf_{\eta \neq {\bf 0}} \frac{\|\Pi(\eta)\|}{\#\tau_{\Pi(\eta)}} >0\nonumber
\,\,\,\Box
\end{equation}

\section{Characterization of SOD as Quasi-Isometry}
\label{s:ISO}
Finally, we obtain a characterization of SOD as quasi-isometry, and, coarse-embedding, respectively.
While the sufficient-part of the proof refers to a previous result~\cite{MoserTSP2014}, see~\ref{AppendixA}, 
the necessity-part is an immediate consequence from Theorem~\ref{th:equDN}.
\begin{corollary}
\label{th:SOD_ISO}
Given $\vartheta>0$, $T\in (0, \infty)$
and the metric spaces $(\mathcal{F}[0, T], \|.\|_{\infty})$ and $(\mathcal{E}_{\vartheta}[0,T], d)$,  
where $d: (\mathcal{E}_{\vartheta}[0,T])^2 \rightarrow [0,\infty)$ is a semi-metric 
induced by the semi-norm $\|.\|$, i.e., $d(\eta_1, \eta_2) = \|\eta_1 - \eta_2\|$ for all $\eta_1,\eta_2 \in \mathcal{E}_{\vartheta}[0,T]$.

Then, the SOD sampling operator $\Phi_{\vartheta}: \mathcal{F}[0, T] \rightarrow \mathcal{E}_{\vartheta}[0,T]$ 
 is a quasi-isometry (coarse-embedding)
if and only if $\|.\|$ satisfies (\ref{eq:ALT}), (\ref{eq:infgeq0}) and 
\begin{equation}
\label{eq:Tinequrho}
\max_{n,m,I}\left\|T^n_{(-+)}\big(T^m_{(+-)}(\eta\big|_I)\big)\right\| \leq \rho(\|\eta\|)
\end{equation}
for all intervals $I \subseteq [0,T]$, $\eta \in \mathcal{E}_1 [0,T]$, where $\rho: [0,\infty) \rightarrow [0,\infty)$ is of the form
$\rho(x) = c\, x + b$, ($c,b>0$) in case of quasi-isometry, and 
$\rho$ is non-decreasing in case of coarse-embedding.
\end{corollary}

Checking the examples of Section~\ref{ss:examples} we obtain the following summary. 

The metric (\ref{eq:M}) satisfies the conditions (\ref{eq:ALT}) and  (\ref{eq:infgeq0}), but not condition (\ref{eq:Tinequrho}), see example
(\ref{eq:MMSN}). The van Rossum metric (\ref{eq:VR}) satisfies the conditions only on a compact interval $[0,T]$.  
Due  to (\ref{eq:EMDMvR}), its multiplicative quasi-isometry constant $A_T$, which equals the discontinuity measure $\Omega_T$, is increasing without limitation if $T\rightarrow \infty$.
Due to (\ref{eq:SchreiberM}), a distance measure induced by a Schreiber similarity measure (\ref{eq:Schreiber}) cannot satisfy the conditions 
 (\ref{eq:ALT})  and (\ref{eq:infgeq0}) simultaneously. For the Victor-Purpura metric we have to distinguish the cases $s=0$ and $s>0$.
In the former case, $s=0$,  (\ref{eq:ALT}) is satisfied, but not (\ref{eq:infgeq0}). 
For a counter example, violating (\ref{eq:infgeq0}), consider two event sequences $\eta_1, \eta_2$ on $[0,T]$, where the first is given by $n$ positive events on $[0,T/2]$ and zero on $(T/2,T]$,  and the second one is given in the opposite way, zero on $[0,T/2]$ and by $n$ positive events on $(T/2,T]$. For the other case, $s>0$, (\ref{eq:VP}) shows, like for the van Rossum metric, that $\Omega_T$ and, therefore, the
multiplicative quasi-isometry constant is increasing without limitation if $T\rightarrow \infty$.

Finally, both, Weyl's discrepancy and the Alexiewicz norm, meet the conditions of Corollary~\ref{th:SOD_ISO} with quasi-isometry constants that are independent from the choice of the time domain $[0,T]$ under consideration. Moreover, 
by replacing the uniform norm $\|.\|_{\infty}$ by the diameter 
\[
\|f\|_{\varnothing} = |\sup_{t\in [0,T]} f(t) - \inf_{t\in [0,T]} f(t)| \leq 2\|f\|_{\infty}
\]
 in the input space, which is a norm on $\mathcal{F}[0,T]$, and choosing either Weyl's discrepancy or
Alexiewicz norm in the sample space, SOD becomes in fact a coarse-isometry, and, further, SOD becomes an asymptotic isometry for 
$\vartheta \rightarrow 0$.  

\section{Conclusion}
In this fundamental research paper we complemented the approach of~\cite{Moser17}.
Above all we presented a full characterization of necessary and sufficient conditions of send-on-delta sampling in order to satisfy the conditions of a quasi-isometric mapping in terms of Hermann Weyl's discrepancy.
Although exemplarily illustrated for SOD, this result also applies to other threshold-based sampling schemes which rely on an evaluation function and 
a non-increasing threshold function.
This result reveals the underlying metric structure of such sampling schemes in terms of quasi-isometry and the fundamental role of
Hermann Weyl's discrepancy metric. 
To this end, the characterization result of this paper entitles the statement that Weyl's discrepancy metric is the canonical metric 
for threshold-based sampling. A direct application of the quasi-isometry property would be a method for evaluating the quality of signal reconstruction. Though the roadmap of this approach has been outlined conceptionally, the details for specific reconstruction techniques, specific classes of signals and noise models remain future research.

\section*{Appendix A}
\label{AppendixA}
A previous result from~\cite{MoserTSP2014} states that
\begin{equation}
\label{eq:PhiNorm}
\|f-g\|_{\varnothing} - 4\vartheta \leq 
\|\Phi_{\vartheta}(f) - \Phi_{\vartheta}(g)\|_D
\leq
\|f-g\|_{\varnothing} + 2\vartheta,
\end{equation}
for all $f \in \mathcal{F}[0,T]$,
where $\|f\|_{\varnothing}= |\sup_{t\in [0,T]} f(t) - \inf_{t\in [0,T]} f(t)| \leq 2\|f\|_{\infty}$ 
denotes the diameter of the graph of $f$, which on $\mathcal{F}[0,T]$ is a norm.

\section*{Acknowledgment}
The author would like to thank the Austrian COMET Program.


\end{document}